\documentclass[prb,twocolumn,aps,shownopacs]{revtex4} 

\usepackage{graphicx}
\usepackage{amsmath}
\usepackage{amssymb}

\begin{document}

\title{\LARGE \textsc{Evanescent states and non-equilibrium in driven superconducting nanowires}}

\author{N. Vercruyssen$^1$, T. G. A. Verhagen$^{1,}$\footnote{Current address: University of Leiden, Leiden, The Netherlands}, M. G. Flokstra$^{2}$, J. P. Pekola$^3$, T. M. Klapwijk$^1$}

\affiliation{(1) Kavli Institute of Nanoscience, Delft University of Technology, 2628CJ Delft, The Netherlands}
\affiliation{(2) Kamerlingh Onnes Laboratory, Leiden Institute of Physics, P.O. Box 9504, 2300 RA Leiden, The Netherlands}
\affiliation{(3) O. V. Lounasmaa / Low Temperature Laboratory, Aalto University, P.O. Box 13500, FI-00076 AALTO, Finland}

\begin{abstract}
We study the non linear response of current transport in a superconducting diffusive nanowire between normal reservoirs. We demonstrate theoretically and experimentally the existence of two different superconducting states appearing when the wire is driven out of equilibrium by an applied bias, called the global and bimodal superconducting states. The different states are identified by using two-probe measurements of the wire, and measurements of the local density of states with tunneling probes. The analysis is performed within the framework of the quasiclassical kinetic equations for diffusive superconductors.
\end{abstract}


\maketitle

\section{Introduction}

Superconducting nanowires are often part of objects to study the Josephson-effects in graphene, carbon nanotubes or semiconducting nanowires. In addition, in many cases superconducting nanowires themselves are used to study their response to radiation. In most cases, the electron back-scattering resistance is assumed to be located at the interfaces and in the normal metal part. An interesting question is to what extent the superconducting mesoscopic or (nano)wires themselves contribute to the resistance of a device due to the conversion from normal current to supercurrent and {\it{vice versa}}. For superconducting nanowires between superconducting contacts, a common assumption is that the applied power leads to dissipation and to an increased temperature varying over the wire length.\cite{Gurevich} In quite a few experiments with a nanowire between normal or superconducting pads, a parabolic temperature profile $T(x)$ is assumed to control the local superconducting properties.\cite{Tinkham2003,Shah,Barends} The definition of a temperature, however, requires that the electrons are in local equilibrium, a condition not easily met for wires of mesoscopic length scales. In the case of a biased normal wire,\cite{Pothier} the diffusion time, $\tau_D={L^2}/D$, with $L$ the wire length and $D$ the diffusion constant, can be much shorter than the inelastic relaxation time $\tau_{in}$. In this case, the electron distribution is highly non-thermal and given by a two-step function $f(E,x)=(1-x)f_0(E-eV/2)+xf_0(E+eV/2)$, with $f_0(E,T)=1/(\exp(E/kT)+1)$ a Fermi-Dirac distribution, $V$ the applied bias, $k$ Boltzmann's constant, $E$ the energy of the electrons measured from the Fermi energy, $T$ the bath temperature and $x$ the coordinate along the wire. 
 \begin{figure}[b!]
 \centering
 \includegraphics[width=\columnwidth]{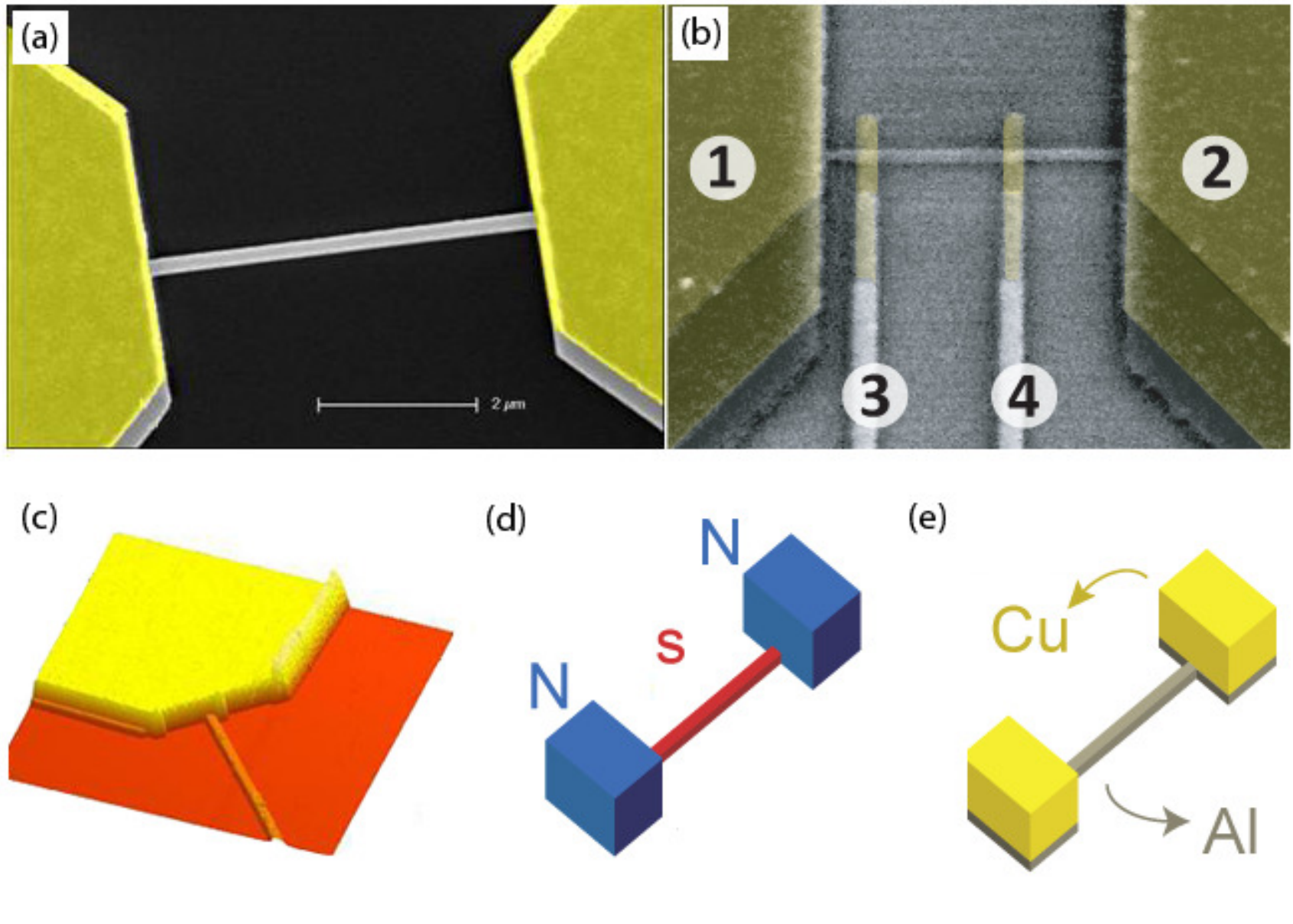}
 \caption{\label{fig:sample} A superconducting Al nanowire connected to two massive normal reservoirs, consisting of the 
 same Al, covered by a normal metal Cu layer: (a) scanning electron microscopy picture, (c) atomic force microscopy picture, (d), (e) schematic representation. The thin Al of the pads is driven normal by the inverse proximity effect of the thick normal Cu. 
 Normal tunneling probes are attached for local measurements (b).}
\end{figure}
A general, non-thermal (or non-equilibrium) electron distribution in a superconductor influences almost all aspects of that superconductor. 
It affects the local Cooper pair density and the current-carrying capacity, but it can also produce a voltage-drop in the superconductor, i.e. a dc resistance of the superconductor. To discuss the various contributions, it is advantageous to separate the non-equilibrium distribution function, $f(E)$, into an energy (or longitudinal) mode, $f_L$, acting primarily on the amplitude of the superconducting gap, 
and a charge (or transverse) mode, $f_T$, which leads to a shift in the pair chemical potential $\mu_{cp}$.\cite{schmid1975} The latter mode $f_T$ describes an imbalance between electron and holes in the excitation-spectrum, leading to a net charge $Q^*$ in the (decaying) excitations. This contribution can be dominant in experiments probing electrical transport in superconducting heterostructures at subpgap energies.

In this paper, we report on an experimental and theoretical study of nonlinear electrical transport in a well-defined model system,\cite{Keizer2006,Sols2001} in which a superconducting wire is connected to two large normal contact pads (Fig. \ref{fig:sample}). The normal electrodes induce evanescent subgap states in the superconducting wire. In addition they act as equilibrium electron reservoirs to fill and empty the states in the superconducting wire. When a bias $eV$ is applied, evanescent electrons and holes are injected from the reservoirs into the superconducting 
wire, and the resulting non-equilibrium distribution function consists of both an energy mode $f_L$ and a charge mode $f_T$. The well defined boundary conditions and simplicity of this system make it a natural choice to study the superconducting state in the presence of a general non-equilibrium.

We address these microscopic properties of the wire experimentally using two point measurements of the nanowire, which are a sensitive probe for the resistive properties originating in $f_T$. Measurements with tunneling probes allow to measure the local density of states and the 
different chemical potentials involved. We demonstrate that two distinct metastable superconducting states exist when the wire 
is driven (Fig. \ref{fig:overview}). The first superconducting state extends over the complete length of the wire, and has been reported in the linear regime by Boogaard et al.\cite{Boogaard2004} The second state exists only under driving, and consists of 
two, geometrically separated superconducting domains, both at the ends of the wire. We show that the superconductivity nucleates in the vicinity of the normal reservoirs, because the local electron distribution is closer to the equilibrium state. The existence of metastable states has been identified in previous work using phenomenological models,\cite{Skocpol1974,Volkov1974} based on a normal resistive domain. We analyze these states using the quasiclassical Green's functions, and show how the energy mode controls the existence of these states, whereas the charge mode controls the resistance. Hence, the full non-linear response is found to be the result of a complex interplay between both the charge and the energy mode non-equilibrium. 

\section{Theoretical framework}
We consider a model system consisting of a superconducting one dimensional diffusive wire connected to two normal, equilibrium reservoirs
(Fig. \ref{fig:sample}d). Electrons are injected into and extracted from the superconducting wire by the reservoirs with equilibrium Fermi 
distributions $f_0(E\pm eV,T)$, with relative Fermi-levels determined by the applied voltage $V$. Within the wire the electrons are distributed over the energies with a position and energy dependent non-equilibrium distribution function 
$f(E,x)$ determined by a diffusion equation. In addition the electronic states concerned are decaying states, evanescent modes, as their energy is smaller than the energy gap ($eV\leq 2\Delta$). Therefore it is necessary to include the interplay between these short-lived states and the 
superconducting condensate, which goes beyond a two-fluid description, in which a sharp distinction between long-living 
quasiparticle states and the condensate is assumed. Such an analysis is performed using the quasiclassical Green's functions theory for superconductivity, which treats the electronic properties of the 
excitations and the condensate on the same footing:\cite{kopnin2001}
\begin{eqnarray}
\check{G}=
\begin{pmatrix}
\hat{G}^R&\hat{G}^K\\0&\hat{G}^A
\end{pmatrix},
\hat{G}^R=
\begin{pmatrix}
G&F_1\\F_2&G^{\dagger}
\end{pmatrix}.
\end{eqnarray}
The retarded (advanced) functions $\hat{G}^{R(A)}$ consist of normal and anomalous propagators $G$ and $F$, which 
describe the single electron spectrum and the coherence between electrons respectively. The occupation numbers of the 
electronic excitations are contained in the Keldysh component $\hat{G}^K$.

In general these Green's functions are dependent on the time, energy, position and momentum of the particle: $G=G(E,t,r,p)$.
However typical variations occur on a much slower length scale than the Fermi wavelength. The Green's functions are 
sharply peaked around the Fermi momentum $p=p_F$, and a considerable simplification can be obtained by integrating $G$ 
over all momenta. A second simplification arises from the short mean free path in dirty superconducting films, which averages 
out any dependence on the momentum direction. The resulting equations were obtained by Usadel\cite{Usadel} and they only 
contain what is called the quasiclassical Green's functions, $g(E,x,t)$ and $f(E,x,t)$. 
Our experimental observations indicate that relevant solutions are stationary, so in addition we neglect all time dependences in the 
equations. This choice is partially supported by theoretical work of Snyman et al.\cite{Snyman2009} who demonstrate for a simplified system that the 
solutions for a dc bias are always stationary. To parametrize $g(E,x)$ and $f(E,x)$ we use a complex pairing angle 
$\theta(E,x)$ describing correlation between electrons and holes, and a complex phase 
$\chi$: $g=\cos(\theta),f_{1,2}=\sin{(\theta)}e^{\pm i\chi}$.\cite{nazarov1996}
The normalization condition $g^2+f_1^\dagger f_2^{ }=1$ is 
automatically fulfilled, while the variations of $\theta(E,x)$ and $\chi(E,x)$ are determined by the following diffusion 
equations:
\begin{eqnarray}
\hbar D\left\{\nabla^2\theta-\sin\theta\cos\theta\left(\nabla\chi\right)^2\right\}\nonumber\\
=-i2E\sin\theta-\cos\theta\left(\Delta e^{-i\chi}+\Delta^* e^{i\chi}\right),\nonumber\\
\hbar D\nabla\left\{\sin^2\theta\left(\nabla\chi\right)\right\}\nonumber\\
=i\sin\theta\left(\Delta e^{-i\chi}-\Delta^* e^{i\chi}\right),
\label{eqn:usaR}
\end{eqnarray}
with $D$ the normal state diffusion constant. The first equation describes how the presence of a local superconducting order 
parameter $\Delta(x)$ generates pair correlations $\theta(E,x)$, which allows to calculate the local density of 
states (DOS) $N(E,x)=\Re\cos(\theta)$. The second equation relates the phase gradient of the gap to the presence of supercurrents.
 
A convenient description of a non-equilibrium superconductor is obtained by introducing a generalized distribution function $h(E)$, 
defined as $G^K=G^Rh(E) - h(E)G^A$. To disentangle the influence of the distribution function on the amplitude and the phase of the order 
parameter, $h(E)$ is split in the even part (energy mode) in particle-hole space $f_L(E,x)$, and the odd part (charge 
mode) $f_T(E,x)$. The total electron distribution functions $f(E,x)$ are then obtained from $2f(E,x)=1-f_L(E,x)-f_T(E,x)$. The presence of a charge mode is related to the presence of a charge $Q^*$ integrated over all excitations, and the consequence of inhomogeneity in the superconducting system, leading to conversion of quasiparticle current to supercurrent. Charge imbalance has been studied thoroughly at temperatures close to $T_c$, {\emph{i.e.}} for long-lived quasiparticle excitations.\cite{Clarke1972,Pethick1979}
However, the concept of charge imbalance also applies to short-lived evanescent states,\cite{Nielsen1982} for small injection voltages and at low temperatures.\cite{Hubler2010,Yagi2006} 

Conservation of energy $E$ and charge $Q$ result in two coupled diffusion equations for $f_L$ and $f_T$:
\begin{equation}
\hbar D\nabla J_E=0,\hbar D\nabla J_Q=2R_Lf_L+2R_Tf_T,
\label{eqn:usaK}
\end{equation}
with
\begin{eqnarray}
J_E&=&\Pi_L\nabla f_L+\Pi_X \nabla f_T+j_{\epsilon}f_T,\nonumber\\
J_Q&=&\Pi_T\nabla f_T-\Pi_X \nabla f_L+j_{\epsilon}f_L\nonumber\\
\Pi_{L,T}&=&1+|\cos\theta|^2\mp|\sin\theta|^2\cosh(2\chi_2),\nonumber\\
\Pi_X&=&-|\sin\theta|^2\sinh(2\chi_2),\nonumber\\
j_{\epsilon}&=&2\Im(\sin^2\theta\nabla\chi),\nonumber\\
R_{L,T}&=&\Re(\sin\theta(\Delta e^{-i\chi}\mp\Delta^* e^{i\chi})),
\label{eqn:usaK2}
\end{eqnarray}
where $\Pi_{L,T,X}$ are generalized diffusion constants, $j_\epsilon$ is the spectral supercurrent and $R_{L,T}$ determine the 
magnitude of the source term on the right hand side of Eq. (\ref{eqn:usaK}).
The energy current is dominated by the diffusion of the energy mode $f_L$.
Our Al wires are relatively short which means we can neglect inelastic processes, as the inelastic  electron-electron and electron-phonon interaction lengths are of the order of 10 $\mu$m at a temperature of 1 K.\cite{Prober1984} For long wires or materials with a strong electron-phonon interaction this is not necessarily true. The stronger electron-phonon coupling of Nb results in a inelastic mean free path of roughly 0.1 $\mu$m.\cite{Kaplan} The charge current consists partly of a normal current driven by a gradient of the charge 
mode, $I_n=\Pi_T\nabla f_T$, and partly of a supercurrent related to a gradient of the phase $I_s=f_Lj_{\epsilon}$. Conversion 
of a normal current into a supercurrent implies a change of $\nabla f_T$, and is proportional to $R_T\approx\Delta$ in Eq. 
(\ref{eqn:usaK}). 

The position dependent potential in the superconductor $e\phi(x)$ is obtained by integrating the charge of 
the quasiparticle excitations over all energies:
\begin{equation}
e\phi(x)=\int_{-\infty}^\infty N(E)f^S_T(E,x){dE}
\label{eqn:pot}
\end{equation}
In order to conserve charge neutrality, the presence of the net charge in the excitations is compensated by a shift in the pair 
chemical potential $\delta\mu_{cp}(x)$. This means that the static electric field $E=\nabla\phi$ which drives the normal current 
$I_n$, does not influence the condensate, since it is exactly balanced by $\delta\mu_{cp}(x)=-e\phi(x)$. If this were not the case, 
the Cooper pairs would accelerate.

The retarded and kinetic equations (\ref{eqn:usaR}, \ref{eqn:usaK}) are completed with the self consistency relation for $\Delta(x)$:
\begin{eqnarray}
\Delta(x)=\frac{N_0V_{eff}}{4i}\int_{-\hbar\omega_D}^{\hbar\omega_D}dE\nonumber\\
\left(\sin\theta e^{i\chi}-\sin\theta^*e^{i\chi^*}\right)f_L
-\left(\sin\theta e^{i\chi}-\sin\theta^*e^{i\chi^*}\right)f_T.
\label{eqn:delta}
\end{eqnarray}
The charge mode is directly related to the observed potential drop over the superconductor through Eq. (\ref{eqn:pot}), 
the energy mode $f_L$ only appears implicitly in the gap Eq. (\ref{eqn:delta}).

\section{Possible Solutions}
\begin{figure}[ht!]
 \centering
 \includegraphics[width=\columnwidth]{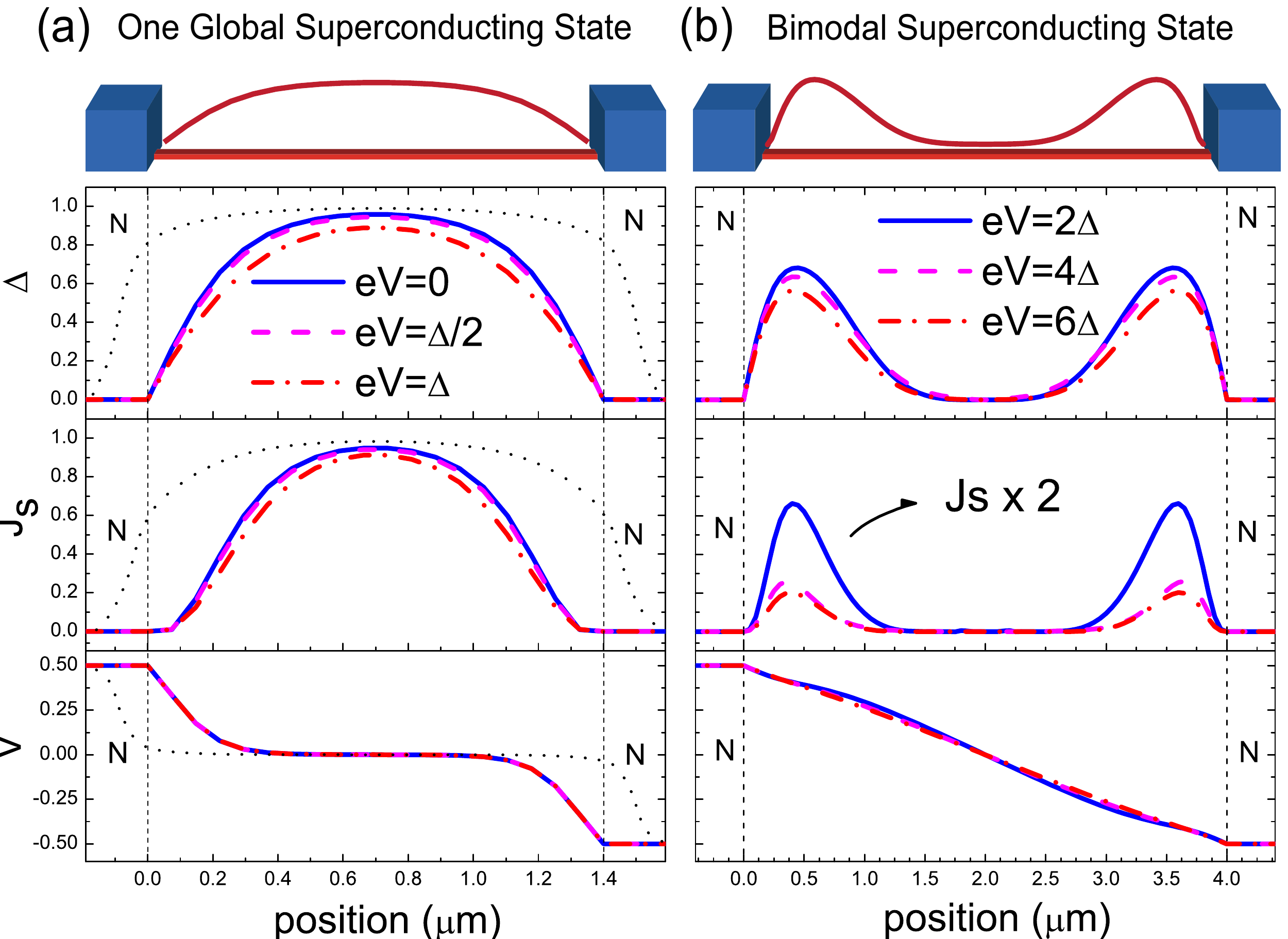}
 \caption{\label{fig:overview} (a) The complete wire is in a single superconducting state with order parameter $\Delta(x)$.
 However near the normal reservoirs the condensate carries only a small fraction $J_s$ of the current as a supercurrent, 
 which results in a resistance and a voltage drop at the ends of the wire, over roughly a coherence length. At the lowest temperatures a small 
 proximity effect can occur at the connection of the bilayer reservoirs to the wire (schematically illustrated by dotted black lines). (b) Two distinct superconducting domains at the ends of the wire are separated by a normal region in the center of the wire. Due to the small supercurrent, the voltage profile is 
 almost equal to the normal state.}
\end{figure}

In this section we present the numerical solutions of Eqs. (\ref{eqn:usaR}) - (\ref{eqn:delta}) for the model system shown in Fig. 
\ref{fig:sample}. The wire can be considered to be one dimensional, as the width and thickness are smaller than the 
dirty superconducting coherence length $w,t\leq\xi=\sqrt{\frac{\hbar D}{2\Delta_0}}$. The normal equilibrium reservoirs act as 
boundary conditions, both for the superconducting pairing-angles $\theta=\nabla\chi=0$ and the distribution functions $f_{L,T}$. 
Temperature enters the problem only through the boundary conditions for $f_L$ and $f_T$, while all non-equilibrium 
processes in the wire itself are contained in the distribution functions. After an initial guess for $\Delta(x)$, the superconducting angles 
$\theta$ and $\chi$ are calculated from the retarded equations (\ref{eqn:usaR}). Subsequently the 
kinetic equations (\ref{eqn:usaK}) can be solved to obtain $f_L$ and $f_T$. Finally the value of 
$\Delta(x)$ is updated using Eq. (\ref{eqn:delta}), and this process is repeated until all values converge. We find two
distinct superconducting solutions for the problem: (a) one global superconducting state (Fig. \ref{fig:overview}a) and
(b) a bimodal superconducting state separated by a normal valley (Fig. \ref{fig:overview}b).

\subsection{One global superconducting state}
\begin{figure}[ht!]
 \centering
 \includegraphics[width=\columnwidth]{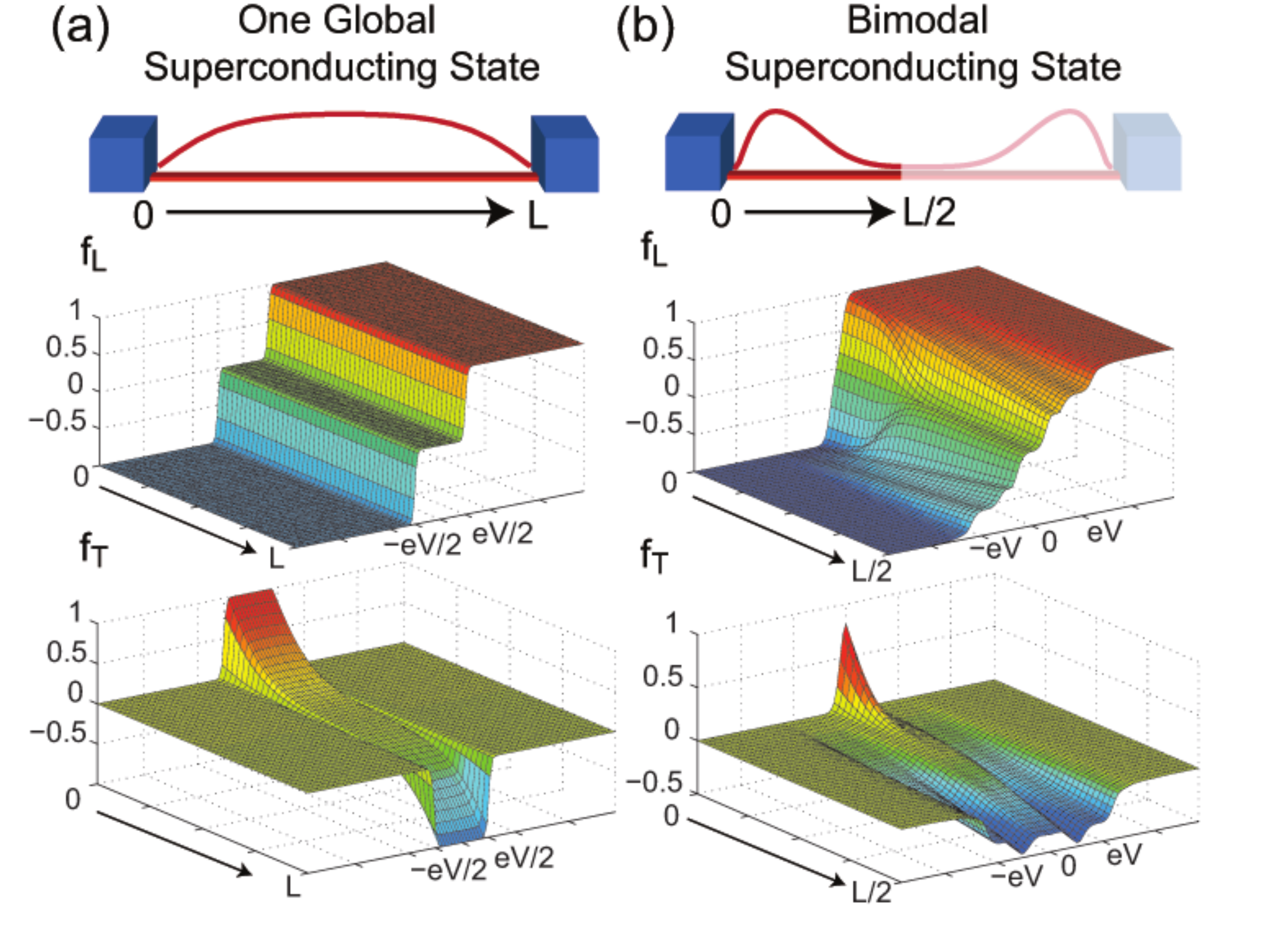}
 \caption{\label{fig:f} The even mode $f_L$ and odd mode $f_T$ of the non-equilibrium distribution function $f(E,x)$. 
 (a) For the global superconducting state, a two-step distribution is present through the full wire, while the charge mode is only present at the edges. (b) A strong, non-thermal energy mode non-equilibrium $f_L$ suppresses superconductivity at the center of the wire.}
\end{figure}

The first solution is characterized by one coherent superconducting state, which extends over the full length of the wire, 
although the strength of the superconducting gap, $\Delta$, is suppressed at the edge of the wire by the presence of the normal 
reservoirs (Fig. \ref{fig:overview}a). Alhough fully superconducting, the wire has a finite resistance due to the conversion 
of a normal current into a supercurrent, as shown by the position-dependent voltage $V$. Normal electrons, which are injected from the metallic 
reservoirs, decay into Cooper pairs over roughly one coherence length $\xi$. The excess charge $Q^*$ associated 
with the charge mode $f_T$ of these evanescent quasiparticle states results in the presence of an electric field in the 
superconductor, and hence a potential drop over the same length the supercurrent increases. These processes correlate with the picture of electrons being injected at energies $E\approx eV$, leading to a two-step distribution $f_L$, as shown previously by Keizer $et$ $al$.\cite{Keizer2006} 
While the charge mode non-equilibrium $f_T$ relaxes over a length scale of $\xi$, because of interaction with the condensate, the energy mode $f_L$ remains constant over the length of the wire due to the absence of inelastic interactions (Fig. \ref{fig:f}a).

 For increasing voltages, there is hardly any change in the profiles of $\Delta,\:\phi,\:J_{s,n},\:f_{L,T}$, until the wire switches to the normal state. For example, there is no gradual expansion of the voltage-carrying parts at the end of the wires, as one would guess qualitatively. A careful analysis \cite{Keizer2006} indicates that the energy mode $f_L$ triggers this transition, while the current is still far below the critical pair breaking current $I_{c0}$. 

In performing these numerical calculations, we assumed that the reservoirs are fully normal down to the lowest temperatures. The dashed lines in Fig. 
\ref{fig:overview}(a) however show a schematic picture of a situation where the reservoirs are proximitized by the wire, which in fact is a situation we encounter in the experiments. The conversion and voltage drop occurs primarily in the contact pads, and the measured resistance is largely a spreading resistance of the contact pad. We will show experimentally that the latter contribution can be quenched by the bias and by a magnetic field. 

\subsection{Bimodal superconducting state}
A second solution was inspired by our experimental results. It consists of two separate superconducting domains located 
at each end of the wire (Fig. \ref{fig:overview}b). A strong energy mode $f_L$ suppresses superconductivity in 
the middle of the wire, while the presence of the cold reservoirs near the ends of the wire favors locally the emergence 
of a gap. Modeling this state is complicated, as the presence of two superconducting regions gives potentially rise to 
time-dependent processes. We can, however, avoid this complication by assuming that the center of the wire is 
fully normal. In that case, it is possible to proceed numerically by splitting the wire in two half-wires and treat them independently, using 
$\theta=\nabla\chi=0$ as boundary conditions. While the distributions at the end of the wire are again given by the 
equilibrium reservoirs, in the middle of the wire we match the distribution functions $f_L$ and $f_T$ and their 
derivatives. The occupation of electronic states with energies $E+eV,E-eV$ are coupled by the applied voltage, 
while previously they were independent. In addition the superconducting potential mixes particle and hole states, 
and one retrieves relatively complex solutions for $f_{L,T}(E,x)$ (Fig. \ref{fig:f}b). At the center of the wire the energy 
mode non-equilibrium is close to a thermal one, but at an elevated temperature similar to a parabolic temperature profile. 
The remaining structure is in essence due to energy-conserving Andreev reflection processes, similar to the electron distribution
in a superconductor-normal metal-superconductor structure.\cite{pierre2001}

\begin{table*}[ht!]
\center
\begin{tabular*}{0.9\textwidth}{@{\extracolsep{\fill}}cccccccccc}
\hline
\#&$L$ ($\mu$m)&$w$ (nm)&$t$ (nm)&$R_n$ ($\Omega$)&$\rho$ ($\mu\Omega$cm)&$D$ (cm$^2$s$^{-1}$)&$T_c$ (K)&$\xi$ (nm)&$R_s$ ($\Omega$)\\
\hline
\hline
1a&1.4&100&90&2.8&1.8&98&1.23&131&1.0\\
1b&2.0&100&90&4.5&2.0&87&1.23&124&0.81\\ 
2&3.0&200&50&3.7&1.23&143&1.35*&152&0.7\\
3a&2.0&100&50&6.2&1.54&115&1.35&135&1.7\\
3b&4.0&100&50&13.3&1.66&106&1.35&131&1.7\\
4&1.5&100&50&5.1&1.70&104&1.35*&129&1.7\\
5&2.0&100&50&4.8&1.20&147&1.35*&154&1.5\\
\hline
\end{tabular*}\caption{Overview of the properties of the different samples: $L$ - length, $w$ - width, $t$ - thickness, $R_n$ - normal state resistance, 
$\rho$ - resistivity, $D$ - diffusion constant, $\xi=\sqrt{\hbar D/2\Delta}$ - coherence length, $T_c$ - critical temperature, $R_s$ - low temperture resistance in the superconducting state. For samples indicated with an asterisk, there is no measurement available for $T_c$. We assumed the same value for $T_c$ as for sample 3 which was fabricated under the same conditions.}
\label{tab:samples}
\end{table*}

The emerging superconducting blobs at the end of the wire are relatively small, both in magnitude $|\Delta|\approx |\Delta_0|/2$ and in
size $L_S\approx4\xi$. Due to their limited size only a tiny fraction of the total current is converted into a supercurrent, and the voltage 
profile is almost identical to the normal state. While the local microscopic properties at the end of the wire show a strong 
superconducting signature, the global properties of the wire are hardly influenced. This is true for the current (which is almost 
completely normal) and the voltage profile, but also for the density of states and the distribution functions. Apart from some small
modifications, the distribution function in the wire is given by a two-step function. The non-equilibrium energy mode $f_L$ is the 
strongest in the center of the wire, and is the main reason why the superconducting state nucleates near the equilibrium 
reservoirs. The influence of $f_T$ is limited as the condensate carries almost no (super)current.

\section{Sample design, fabrication and characterization}

Figure \ref{fig:sample}(a) shows a typical superconducting Al nanowire contacted by two massive normal reservoirs, 
consisting of the same thin Al layer covered by a thick Cu layer. For reasonably clean interfaces the inverse proximity 
effect of the thick Cu drives the Al normal down to the lowest temperatures. The massive volume of the contacts 
guarantees that they act as equilibrium reservoirs from which electrons are injected into the wire. When a bias is
applied however, the temperature of the reservoirs electron distribution function $f_0(E,T)$ might for increasing voltage deviate from the bath 
temperature according to:\cite{Henny1999}
\begin{eqnarray}
T^2&=&{T_0^2+b^2V^2},\\
b^2&=&\frac{1}{\pi L}\frac{R_\square}{R_{wire}}\ln\left(\frac{r_0}{r_1}\right),
\label{eqn:tbath}
\end{eqnarray}
where $L$ is the Lorenz number, $R_{\square}$ the sheet resistance of the contact, and $r_0$ and $r_1$ respectively 
the electron-electron and electron-phonon inelastic mean free path. 
The temperature increase can be considerable, 
and the most obvious way of decreasing it is to minimize the ratio $R_\square/R_{wire}$ by using thick reservoirs, which we have implemented in our sample design.

The samples are realized by three angle shadow evaporation through a suspended resist mask (PMMA/LOR double 
layer), in a system with a base pressure of 0.5 - 1.5 x 10$^{-7}$ mbar. The parameters of the different samples are summarized in Table \ref{tab:samples}. First 50 - 90 nm of 99.999\% purity Al is deposited through a slit in the suspended mask to create the superconducting wire and the thin bottom layer of the pads. Evaporation of a thick (200 - 500 nm) copper layers under an angle, which avoids deposition through the slit, completes the normal bilayers forming the reservoirs. The time between the two steps is kept to a minimum ($<$ 10 min) to ensure a clean and transparent interface. Subsequently the Al is oxidized during 5 minutes in a pure O$_2$ atmosphere with a pressure of 4.6 mbar to create an AlO$_x$ tunnel barrier of $R_nA\approx300\:\Omega\mu$m$^2$. The Cu probes are deposited during the last evaporation step under a second angle. The size of the wires is measured using scanning electron microscopy. The thickness was obtained from a quartz crystal monitor used during the deposition of the Al film, and calibrated by atomic force microscopy. 

\subsection{Linear response of the nanowire}
\label{sec:linear}

\begin{figure}[!h]
 \centering
 \includegraphics[width=\columnwidth]{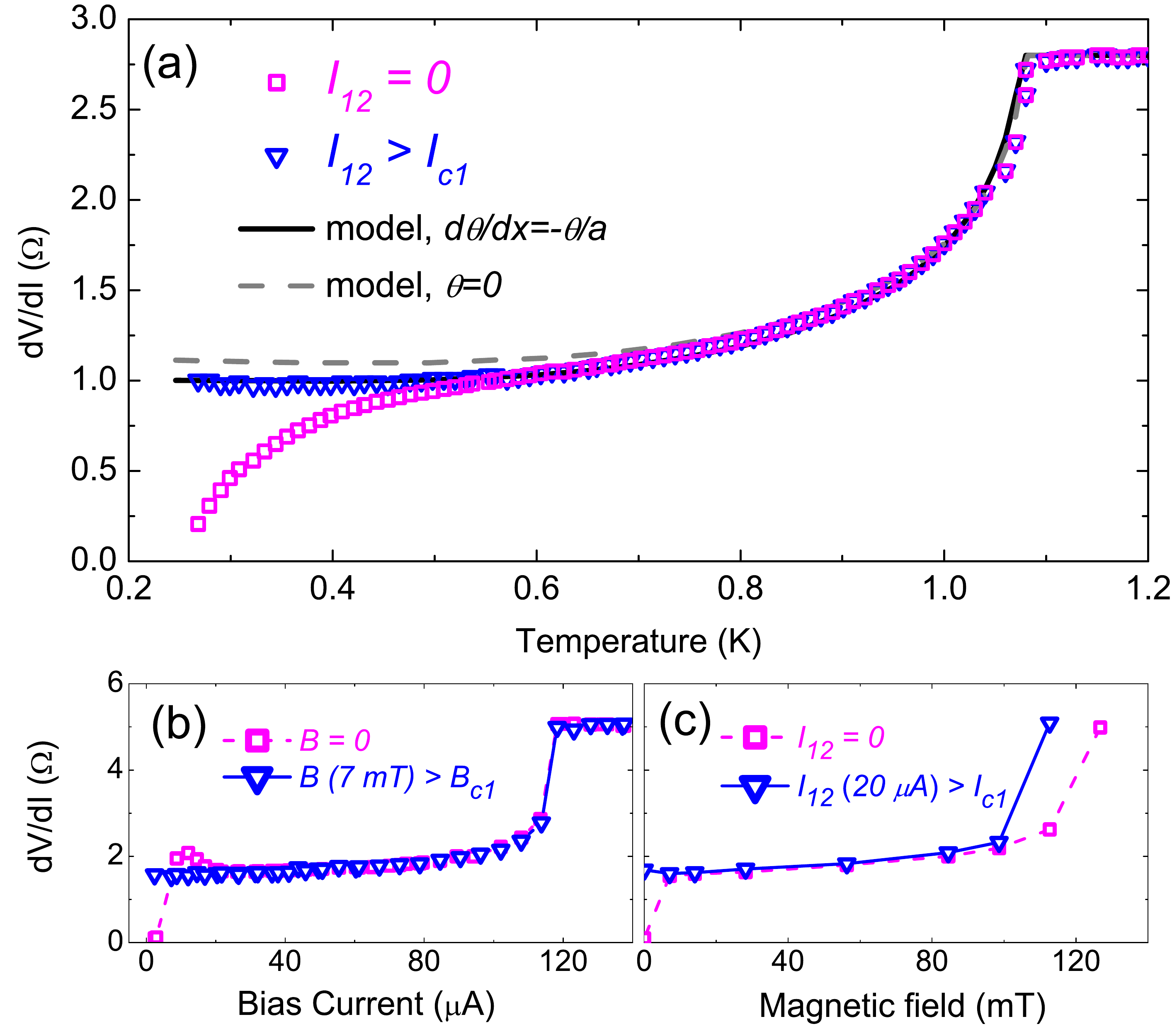}
 \caption{\label{fig:RT} (a) The two probe resistance versus temperature of a 1.4 $\mu$m long wire (sample 1a).
 Due to the proximity effect of the wire on the normal reservoirs, the resistance becomes negligible at low temperatures. 
 This weak proximity effect can be suppressed by applying a small bias current (b) or small magnetic field (c) (sample 4, 200 mK). This 'corrected' wire resistance is constant down to the lowest temperatures (magenta squares of panel a). A model (dashed line) with rigid normal boundary conditions for the pairing angle $\theta=0$ slightly overestimates the observations. A weaker boundary condition (full line), in which $\theta$ decays gradually to zero over a characteristic length $a$ shows excellent agreement with the experiment. }
\end{figure}

\begin{figure*}[ht!]
 \centering
 \includegraphics[width=\textwidth]{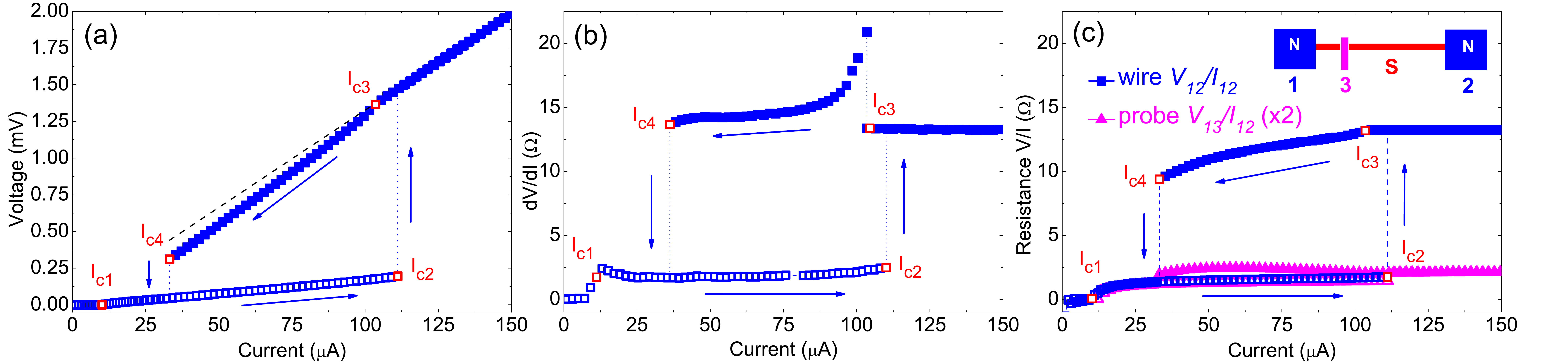}
 \caption{\label{fig:global}The voltage $V_{12}$ (a) and differential resistance (b) of a 4 $\mu$m long wire (sample 3b) 
 as a function of bias current $I_{12}$, measured at 200 mK. We define four different regimes with boundaries labeled $I_{c1}-I_{c4}$, 
 each characterized by a nearly constant differential resistance. The critical currents $I_{c2}$ and $I_{c4}$ are 
 defined as the bias currents where the wire switches between the two hysteretic voltage branches.  $I_{c1}$ and $I_{c3}$ are the transition points between the two different states of one branch. (c) The apparent resistance of the complete wire $V_{12}/I_{12}$, and of the edge of the wire $V_{13}/I_{12}$ as measured with a voltage probe, multiplied by two for the ease of comparison.}
\end{figure*}

Figure \ref{fig:global}a shows a typical current voltage curve (IV). The linear regime extends up to a critical current designated by $I_{c1}$. This initial slope has been measured as a function of temperature with an ac technique leading to the results shown in Fig. \ref{fig:RT}. We used a bias current $I_{12}$ of 1 $\mu$A modulated at 342 Hz (terminals labels are shown in Fig. \ref{fig:sample}b). The two point resistance of this 1.4 $\mu$m long wire (sample 1a) as a function of temperature displays a well-defined pattern (open squares). The spreading resistance of the contact pads adds a small but finite contribution of approximately 20 m$\Omega$ to the measured two point resistance. 
Clearly, at high temperatures the wire is normal and has a resistance $R_n$. When the temperature is decreased below $T_{c}=1.05$ K the resistance of the wire drops considerably as it becomes superconducting. This critical temperature is depressed compared to the intrinsic critical temperature of the aluminum due to the proximity-effect, as discussed by Boogaard et al.\cite{Boogaard2004} For intermediate temperatures (500-800 mK) the resistance appears to saturate at a value $Rs\approx1\ \Omega$. As we will analyze further, this is the result of a normal current penetrating into the wire over roughly one coherence length $\xi$, yielding a $R_{s}\approx 2\rho\xi/A$, with $A$ the crosssection. 

Further lowering of the temperatures leads to a further drop in resistance to almost zero, suggesting that the bilayer contacts are becoming superconducting, due to a low transparency of the interface between the Al and Cu layers. To check this hypothesis we measured the resistance of identical Al/Cu bilayer strips down to the lowest temperatures, 
and find that they stay normal. Instead we attribute the vanishing resistance due to the proximity effect by the nanowire on the contact point between the normal reservoir and the superconducting wire itself. The superconductivity gradually spreads out into the bilayer, leading to a normal-superconducting boundary which moves from the nanowire into the contact pads (Fig. \ref{fig:overview}a). Consequently, the current conversion resistance in the wire itself becomes gradually less relevant. As the cross-section for the conversion moves into the contacts it becomes larger, reducing its resistive contribution. Hence, only a part of the spreading resistance ($\approx$ 20 m$\Omega$) is measured. 

This observed pattern changes if we measure the resistance for a small dc bias current, larger than $I_{c1}$. 
Then, superconductivity in the weakly proximitized region in the pads is suppressed, and the differential resistance stays constant down to the 
lowest temperatures (blue triangles Fig. \ref{fig:RT}a). 
To further test this hypothesis we measured the IV of the wire while we apply a small magnetic field of 7 mT, parallel to the pads 
but perpendicular to the wire (Fig. \ref{fig:RT}b). The vanishing resistance at zero bias is no longer observed, while the differential 
resistance at higher biases is identical to the one without magnetic field. This indicates that such a small field does not influence the 
properties of the wire, and only quenches the weakly proximitized region in the pads. Only at a much higher field $B\approx$ 100 mT 
do we observe a change in the differential resistance of the wire (Fig. \ref{fig:RT}c). 

The dashed gray line in Fig. \ref{fig:RT} shows the calculated two-point resistance. The bulk critical temperature $T_{c0}=1.23\ K$ 
was the only free parameter in the fit, while the diffusion constant $D=98$ cm$^2$/s was obtained through the relation $D=\rho/N_0e^2$. 
The resistivity $\rho$ is deduced from the normal state resistance $R_n$, using $N_0=2.2\cdot10^{47}$ J$^{-1}$m$^{-3}$ for the density of states at the Fermi level.\cite{ashcroft} The superconducting coherence length 
is obtained from $\xi=\sqrt{\frac{\hbar D}{2\Delta}}$. Although the numerical calculation agrees quite well with the data (for $I>I_{c1}$), 
the model overestimates the residual resistance at low temperatures. This indicates that the assumption of completely normal
contact pads is too rigid, as also observed by Boogaard et al.\cite{Boogaard2004} To include the geometric out-diffusion 
of coherent electrons into the normal pads, we adjust the boundary conditions at the ends of the wire to: $\nabla\theta=-\theta/a$, which indicates the dilution of superconductivity into the normal pads over a characteristic length scale $a$. With $a\approx18$ nm (full line in Fig. \ref{fig:RT}) we find excellent agreement with the observations. The key parameters are listed in Table \ref{tab:samples} for the different samples. 
It demonstrates that the linear response of the wires is well understood, but that the boundary conditions are a sensitive part of the problem even for the thick and wide contact pads used. However for bias currents $I\geq I_{c1}$ the system is in a well defined state, which can be connected to the theoretical predictions.

\subsection{Characterization of the tunnel probe}

To measure locally the density of states, the electrostatic potential $e\phi(x)$ and the chemical potential of the condensate $\mu_{cp}$ we use a normal tunneling probe. The current flowing from a normal tunnel probe contacted to a non-equilibrium superconductor at a position $x$ is given by:
\begin{eqnarray}
I_T(V,x)=\frac{1}{eR_n}\int_{-\infty}^\infty\Re\{\cos(\theta(E,x))\}\nonumber\\
\{f_T^S(E,x)-f_T^N(E+eV)\}dE,
\end{eqnarray}
with $\theta(E,x)$ the pairing angle, $f_T^S(E,x)$ the charge mode non-equilibrium distribution in the superconductor, and $f_T^N=1-f_0(E+eV)-f_0(-E+eV)$ the distribution function in the normal probe. Using Eq. (\ref{eqn:pot}) we can rewrite this to:
\begin{equation}
I_T(x)=\frac{1}{eR_n}\{e\phi(x)-\int_{-\infty}^\infty{N(E,x)f_T^NdE}\}.
\label{eqn:tj}
\end{equation}
The tunnel current consists of two contributions, the first one does not depend on the applied voltage, but is completely 
determined by the charge imbalance in the superconductor, leading to the local electrostatic potential $\phi(x)$. The second contribution is given by the convolution of the local DOS $N(E,x)$ of the superconductor and the distribution function $f^N_T(E+eV)$ in the normal metal. At low temperatures the differential conductance of the tunnel contact is a direct measure for the density of states in the superconductor. 
The condensate chemical potential of the superconductor can be 
obtained from $N(E,x)$, which is symmetric around $E=\mu_{cp}$. 
\begin{figure}[ht!]
 \centering
 \includegraphics[width=\columnwidth]{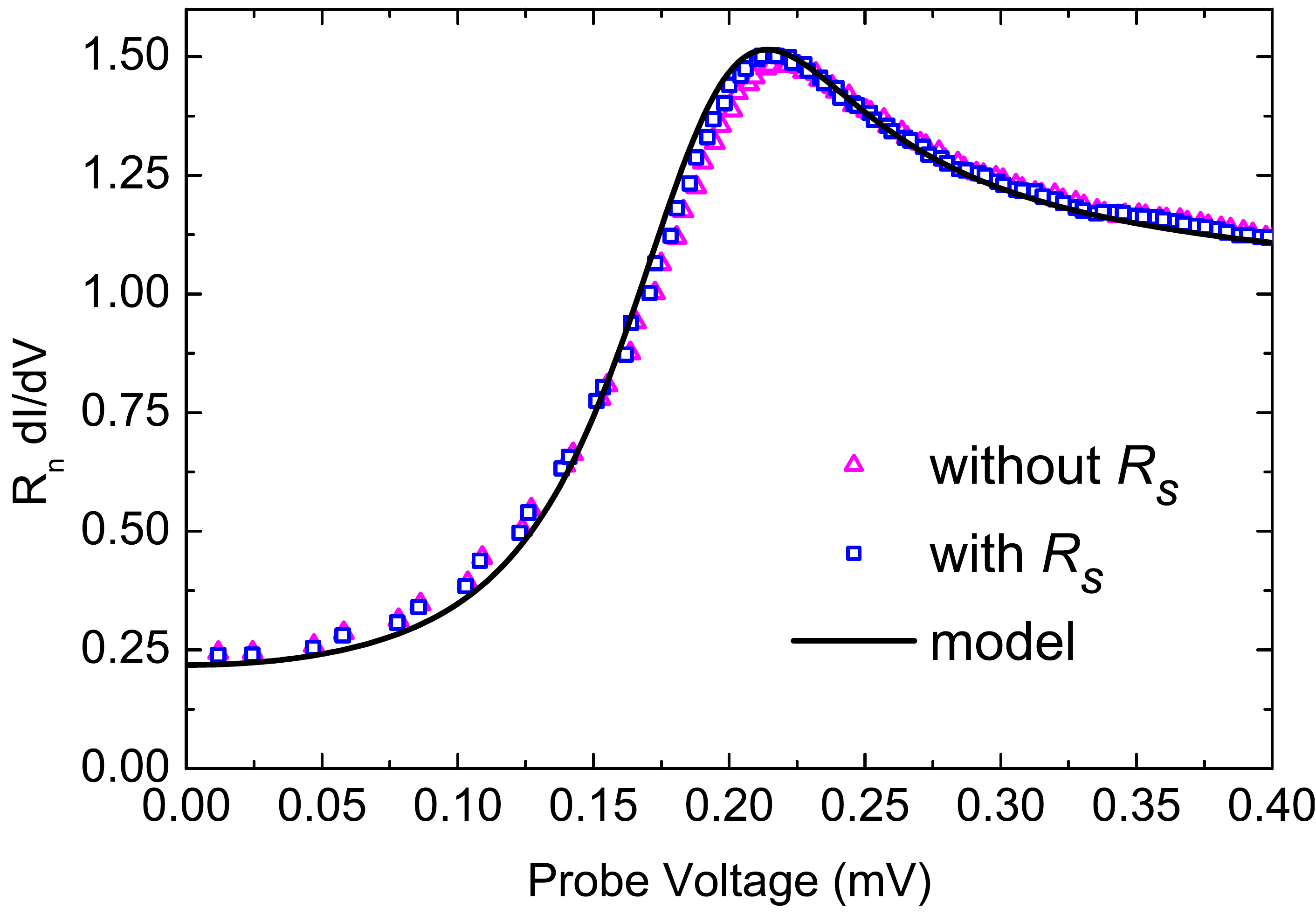}
 \caption{\label{fig:dosUPa}The differential conductance of the local tunnel probe as a function of applied voltage (magenta triangles).
 Good agreement between experiment and theory (full line) is obtained when the series resistance of the set-up is included (blue squares).}
\end{figure}

Figure \ref{fig:dosUPa} shows a typical measurement of the differential resistance for a tunnel probe located at a distance of 320 $nm$ $=2.4 \xi$ from the normal reservoir of a 4-$\mu$m-long wire. The nanowire is biased just above $I_{c1}$, to ensure it is in a well-defined state. 
The bias current needed to drive the probe is typically four orders of magnitude smaller than the bias current of the nanowire, due to the high normal resistance of the tunnel junction ($R_T=43$ k$\Omega$). Hence, it is safe to assume that the properties of the nanowire are not influenced by the measurement of the probe.
One recognizes the coherence peaks at the gap voltage, however, the subgap DOS is increased in comparison to the BCS values due to the presence of the normal banks, and the driving of the nanowire. The simulated local DOS, for the set of parameters, is in good agreement with the data, but near the gap voltage a small discrepancy exists.
We attribute this to a series resistance in the wiring of the tunnel probe and can correct our data for this contribution. We obtain a good agreement 
between the data and the theory using a series resistance of $R_S=1.2$ k$\Omega$, which is the estimated wiring resistance of the experimental set-up.

\section{Two-state analysis and discussion}

We have realized and studied a total of seven samples with parameters shown in Table \ref{tab:samples}. All displayed similar behavior. The non linear current voltage characteristic of a typical sample is shown in Fig. \ref{fig:global}(a), with two clearly distinguished branches. Before discussing the details we first indicate the various signatures for processes, which dominate the various regimes. 
Increasing the current from zero bias we pass $I_{c1}$ the current at which the proximitization in the banks is quenched as discussed in Section IV.A. Beyond $I_{c1}$ until $I_{c2}$, we claim that the wire remains in the global superconducting state, characterized by a low and almost constant differential resistance $R_s$ (Fig. \ref{fig:global}b). This resistance 
reflects the conversion of a normal current into a supercurrent and is located at the edges of the wire. At the current $I=I_{c2}$ 
the wire switches into the normal state, leading to an abrupt switch of both the voltage and the differential resistance, followed by a constant differential resistance equal to the normal state resistance. 

Decreasing the current from the normal state, a kink in the measured voltage signals a more subtle transition at $I=I_{c3}$. The measured voltage shows a small deficit with respect to its normal state value (black dashed line) suggesting the nucleation of superconductivity. We claim that superconductivity nucleates here at the ends of the wires close to the contact pads in agreement with Fig. \ref{fig:overview}b. The sudden transition at $I=I_{c4}$ is due to the transition from the bimodal to the global superconducting state.

A first experimental indication to support this interpretation is provided by Fig. \ref{fig:global}c, which compares local measurements with measurements over the full wire. It shows the two-point resistance of the wire $V_{12}/I_{12}$ (squares) as a function of bias current $I_{12}$, 
together with the apparent resistance $V_{13}/I_{12}$ (triangles) at the ends of the wire (see inset for the probe-position and terminal labels). 
The probe voltage is multiplied by two for comparison, as a similar contribution is present at the other edge of the wire.
For the lower branch, the assumed global superconducting state, one observes that the voltage drop $V_{13}$ over the end of the wire is almost identical to half of the complete voltage drop over the wire-length, a direct proof that this resistance is located at the ends of the wire.

In contrast, in the normal state, the voltages $V_{12}$ and $V_{13}$ are, as expected, proportional to their respective lengths along the wire, $V=\rho L/A$. Upon decreasing the bias below $I_{c3}$, where we assume the bimodal state exists, one observes over the full length of the wire a 
decreasing resistance for decreasing bias, signaling the growing strength of superconductivity somewhere. The measured resistance over 
the end of the wire, however, increases compared to the normal state $V_{13}/I_{12}\geq R_n$. Though counter-intuitive 
this is consistent with the general non-equilibrium present in the superconductor. In the following we make a detailed analysis of both 
superconducting states, and place the experimental results in the context of the theoretical model. 

\subsection{Global superconducting state}

\begin{figure}[ht!]
 \centering
 \includegraphics[width=\columnwidth]{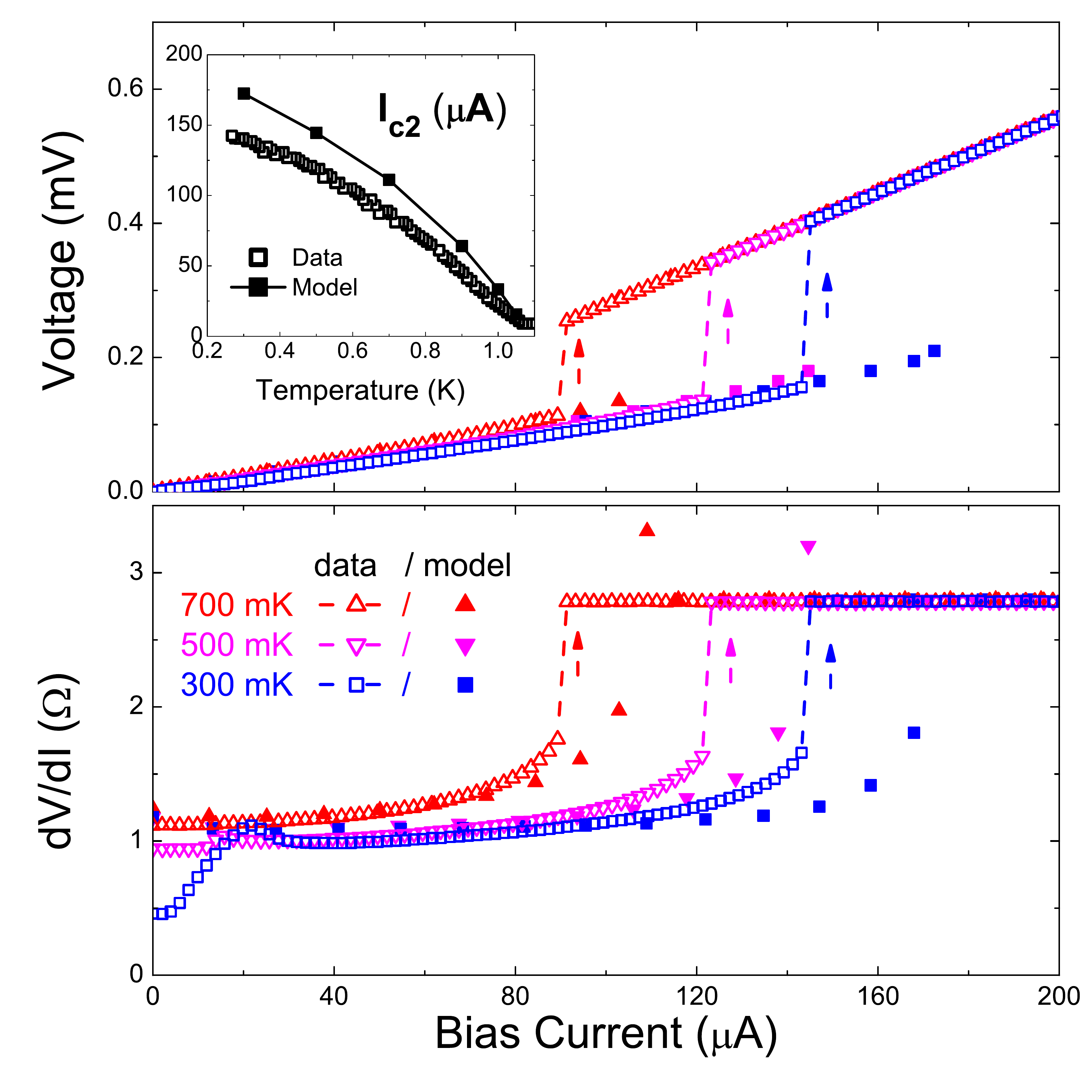}
 \caption{\label{fig:nsnIV} Two probe voltage (a) and differential resistance (b) as a function of bias current, for a 1.4 $\mu$m wire, 
 at three different bath temperatures. Open symbols: experimental data. Filled symbols: numerical simulations. The critical current 
 as a function of temperature (inset).}
\end{figure}

Figure \ref{fig:nsnIV} shows two-point measurements of the lower branch of a 1.4 $\mu$m long nanowire (Sample 1a) at three different bath
temperatures. In view of the analysis shown Section \ref{sec:linear} we assume that the resistance of the wire is primarily determined by the charge mode of the distribution function $f_T(E,x)$,
which depends on the position dependent density of states and the order parameter $\Delta(x)$. The weak dependence of the 
differential resistance on the current indicates that the superconducting properties of the wire hardly change with increasing bias (open symbols).
Although numerical simulations (filled symbols in Fig. \ref{fig:nsnIV}) show the same qualitative behavior, the simulations seem 
to overestimate the bias current at which the differential resistance begins to increase. Hence, the observed switching
current $I_{c2}$ is also slightly lower than predicted. Ignoring this small discrepancy, the simulated data show good 
agreement with the experiment over the complete temperature range (inset Fig. \ref{fig:nsnIV}). At the same time, the observed values 
for the critical current (or critical voltage) are much smaller than what one would expect for a pair-breaking current, experimentally,\cite{Romijn1982,Anthore2003} as well as theoretically.\cite{Lukichev1980,Bardeen1962} This demonstrates that the non-equilibrium processes should be taken into account in evaluating the parameters. 
The remaining deviations between theory and experiment 
suggest, most likely, that the temperature of the reservoirs deviates from the bath temperature for higher driving currents, 
as expected from Eq. (\ref{eqn:tbath}).
 
Figure \ref{fig:dos}(a) shows the differential conductance $dI_{13}/dV_{13}$ of a tunnel probe, located at a distance of 320 nm (=2.4 $\xi$) 
from the normal reservoir of a 4 $\mu$m long wire (Sample 3b). The wire is biased at a fixed current $I_{12}$ with a corresponding voltage $V_{12}$. 
At the same time the probe current $I_{13}$ is varied while measuring the probe voltage $V_{13}$ (Fig. \ref{fig:sample}b).
The evolution of the local DOS for increasing bias is shown on the right side in Fig. \ref{fig:dos}a. The conductance at zero bias $V_p=0$ increases 
slightly for increasing bias, while the coherence peaks get further smeared out. The dependences are however weak, and even right 
before the switching current $I_{c2}$ the DOS is hardly affected by the drive current. These observations are in
close agreement with the theoretical predictions and confirm the idea that the superconducting state remains globally stable. For increasing bias the resistance remains located at the ends of the wire and the DOS does not change either. Unfortunately, we have not been able to directly measure $f_L(x)$. Nevertheless we believe that this energy-mode non-equilibrium triggers the transition at $I_{c2}$ as analyzed by Keizer et al.\cite{Keizer2006} 

\subsection{Bimodal superconducting state}

\begin{figure}[ht!]
 \centering
 \includegraphics[width=\columnwidth]{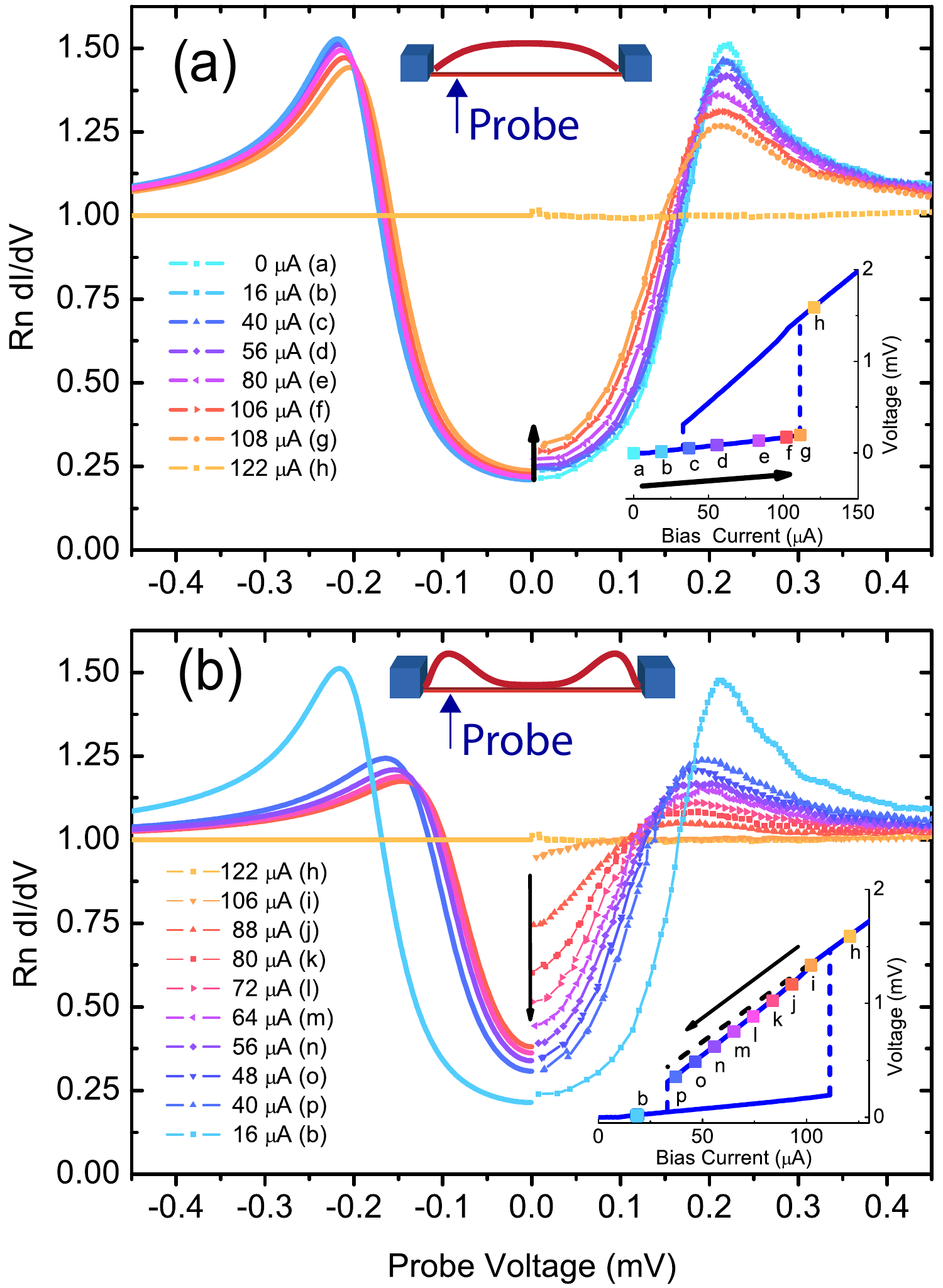}
 \caption{\label{fig:dos}The local density of states for (a) the global superconducting state and (b) the bimodal state, for different bias currents $I_{12}$ of the nanowire, measured at 200 mK. For the global superconducting state, the gap is only weakly dependent on the bias current, while for the bimodal state, one observes a DOS gradually changing from a normal into a superconducting state.}
\end{figure}

The continuous transition, with decreasing bias, from the normal into a superconducting state at $I_{c3}$ (Fig. \ref{fig:global}) indicates that the emerging superconducting state is initially very close to the normal state. For lower bias currents, the absolute resistance gradually decreases 
(Fig. \ref{fig:global}c), which suggests that an increasing fraction of the current is carried by the emerging condensate (Fig. \ref{fig:overview}b). 
 A similar picture is observed for the local density of states, plotted in Fig. \ref{fig:dos}b for different bias currents $I_{12}$ 
 of the nanowire. Below $I_{c3}$, a gradually increasing gap is found, unambiguously showing the emergence of superconducting 
 order. Close to $I_{c3}$, the DOS at the position of the probe evolves in a continuous way from a flat spectrum into a spectrum with a gap. However, at $I_{c4}$ one observes 
 an abrupt transition to a situation with a stronger gap. The abruptness indicates that it is a transition from two distinct superconducting states, which directly proves that at least two microscopically distinct superconducting states exist. Although the simulations for the local DOS
 agree well at currents close to $I_{c4}$, they do not account in detail for the gradual evolution between the normal and superconducting state at $I_{c3}$. At this point, we assume that the reservoirs start to heat up, and can no longer be treated as equilibrium reservoirs with $T=T_0$. Overall, the model supports the picture of the emergence of the superconducting state quite nicely, with the strongest non-equilibrium in the wire occurring at $I_{c4}$, with the reservoirs most closely to equilibrium at $T=T_0$.

The electro-chemical potential of the superconducting condensate, $\mu_{cp}$, is determined from the minimum of the measured 
DOS $\mu_{cp}=eV|_{\min(DOS)}$. It is found that at this probe-position $\mu_{cp}$ is equal to the electrostatic potential $V_1$ of the adjacent reservoir. 
Measurements with a probe in the middle of the wire show that, in the same bias regime, the voltage is equal for both sides of the wire, which means that the state is symmetric, and a similar superconducting region should exists near the other reservoir at a potential $V_2$. 
If these two regions were part of one global superconducting state, there would be a voltage drop $\Delta V=V_1-V_2$ over the superconducting potential $\mu_{cp}$ of this state, and a superconducting phase-slip process should occur. However, according to the Josephson relation $2eV=\partial\chi/\partial t$ it would be at a frequency $\nu\approx8\hbar/\Delta$, which is too high compared to the energy-gap. 

The fact that the two-point resistance is so close to $R_n$, the gradual increase of the DOS at the position of the probe, and the electro-chemical potential of the condensate demonstrate that two separate superconducting regions emerge at the edges of the wire. The physical reason is the energy 
mode non-equilibrium, as discussed by Keizer et al\cite{Keizer2006} for the lower branch, but similarly for this upper branch. At the bias $I_{c4}$ the wire is still largely normal and $f_E$ is given by the two-step distribution-function. In the middle of the wire the width of the step is several times bigger than the superconducting gap. Through relation Eq. (\ref{eqn:delta}) it is seen that this suppresses fully the nucleation of a gap, while the cold equilibrium 
reservoirs favor the emergence of a gap at the edges of the wire. Simply put, the ends are cold where the 
center of the wire is hot.Therefore we conclude that the results are most easily understood as due to two distinct 
superconducting domains, separated by a normal central region, what we have called the bimodal state. 

Finally, we discuss the voltage $V_{13}$ measured by the probe when the wire is biased into the bimodal state (triangles Fig. \ref{fig:global}c). 
Close to $T_c$ the voltage measured by such a normal probe is equal to the electrochemical potential of the quasiparticle bath.\cite{Pethick1979} 
At low temperatures\cite{Clarke1972} and for short-lived quasiparticles\cite{Nielsen1982} it is impossible to define a quasiparticle bath with a well-defined chemical potential, however the measured voltage is still related to the local electrostatic potential $e\phi$ (using Eq. \ref{eqn:tj}):
\begin{eqnarray}
e\phi=\int_{-\infty}^\infty{N(E)f_T^N(E+eV)dE}.
\end{eqnarray}
For a relatively small charge imbalance $e\phi(x)\ll\Delta$, the measured voltage equals the local electrostatic potential $e\phi(x)$ divided by the local DOS in the superconductor at zero energy: $V\approx \phi/N(0)$. Hence the voltage measured with the tunnel probe can be larger than the local potential $e\phi(x)$.
 
\section{Conclusion}
We have analyzed a well-defined model-system of a superconducting wire between two massive normal contact pads. We demonstrate that this system, when driven by a current, has two distinct metastable superconducting states. 

For low bias we find a global superconducting state with most of the resistance occurring as a current-conversion resistance at the ends of the superconducting wires where normal current enters. Although resistive, we demonstrate that the whole wire including the edges continues to be in one coherent superconducting state. This state does hardly change for increasing current, until the wire switches abruptly to the full normal state at a current, 
which is much lower than the critical pair-breaking current. On a microscopic level, the distribution function changes considerably 
and is strongly different from the commonly used parabolic temperature profile. A numerical analysis based on the non-equilibrium quasiclassical 
Green's functions shows that the switching current is determined by the non-equilibrium electron distributions, in good agreement with 
the experimental results.

For high bias, decreasing the current from a fully normal state, we find that the superconducting state emerges as two decoupled 
domains at the ends of the wire. The vicinity of the cool equilibrium reservoirs favors the nucleation of the superconducting state at these ends, while strong non-equilibrium at the center of the wire continues to suppress the superconductivity. Upon further lowering of the 
bias current, the two domains grow in strength until the wire switches back to the low resistive, globally superconducting state. We speculate, that the transition from one state to the other, is triggered by a condition in which the Josephson coupling energy between the two domains exceeds the thermal energy at that bias point. 

This work is also relevant for normal metal-superconductor-normal metal mixing devices, called hot-electron bolometer (HEB) mixers.\cite{Prober1993} In most practical cases, the superconducting material is thin NbN and gold (Au) normal pads are used as antenna. 
Under the condition that no radiation is applied to an HEB, the present analysis is helpful to understand the observed current-voltage characteristics, which are analogous to the one shown in Fig. \ref{fig:global}a.\cite{Hajenius}
The resistive properties for low bias and temperature will be dominated by the conversion resistance at the interfaces (controlled by $f_T$). This regime will extend to a critical current, analogous to $I_{c2}$ reported here, but with a value which may depend on the electron-phonon relaxation, which is present in a material like NbN, but is negligible in our experiment with Al. Beyond this critical value, the device is most likely either fully in the normal state (beyond $I_{c3}$ as identified here), or in the bimodal state (for lower biases between  $I_{c3}$ and  $I_{c4}$). The stronger electron-electron and electron-phonon interaction in NbN as compared to Al, will bring the longitudinal non-equilibrium, $f_L$, closer to a local thermal profile. In case radiation is applied to an HEB, an overall increase in electron temperature occurs, which brings the superconductor close to its transition point where thermally activated phase slip events contribute to the resistivity. Hence, for a full understanding of the HEB mixers, one needs to take into account two contributions to the observed resistance: first the static conversion resistance inside the superconductor near the interface between the normal metal and the superconductor, described here, dominating for the unexposed devices, and second the resistance due to time-dependent phase-slip events occurring at electron temperatures close to the critical temperature of the superconductor, which dominates under actual mixer operation.\cite{Barends} 

We would like to acknowledge Nanofridge (Nanosci-ERA), Microkelvin (No. 228464, Capacities Specific Programme) and the Foundation for fundamental research on matter (FOM).

\end{document}